\documentclass[twocolumn,pre,showpacs,eqsecnum,floatfix]{revtex4}

\usepackage{dcolumn}
\usepackage{amsfonts}
\usepackage{amsmath}
\usepackage{amssymb}
\usepackage{bm}
\usepackage{graphicx}
\usepackage[usenames]{color}

\begin{document}

\newcommand{\AWchange}[1]{{\color{red}{#1}}}
\newcommand{\AWcomment}[1]{{\color{PineGreen}{#1}}}
\newcommand{\SLcomment}[1]{{\color{ProcessBlue}{#1}}}
\newcommand{\SLchange}[1]{{\color{BurntOrange}{#1}}}
\newcommand{\brm}[1]{\bm{{\rm #1}}}
\newcommand{\tens}[1]{\underline{\underline{#1}}}
\newcommand{\mm}{\overset{\leftrightarrow}{m}}
\newcommand{\xv}{\bm{{\rm x}}}
\newcommand{\Rv}{\bm{{\rm R}}}
\newcommand{\uv}{\bm{{\rm u}}}
\newcommand{\nv}{\bm{{\rm n}}}
\newcommand{\Nv}{\bm{{\rm k}}}
\newcommand{\ev}{\bm{{\rm e}}}
\newcommand{\ar}{r}
\newcommand{\scp}{a_t}

\title{Smectic-$C$ tilt under shear in smectic-$A$ elastomers}

\author{Olaf Stenull}
\affiliation{Department of Physics and Astronomy, University of
Pennsylvania, Philadelphia, PA 19104, USA}

\author{T. C. Lubensky}
\affiliation{Department of Physics and Astronomy, University of
Pennsylvania, Philadelphia, PA 19104, USA}

\author{J. M. Adams}
\affiliation{Cavendish Laboratory, JJ Thomson Avenue, Cambridge CB3
0HE, United Kingdom}

\author{Mark Warner}
\affiliation{Cavendish Laboratory, JJ Thomson Avenue, Cambridge CB3
0HE, United Kingdom}

\vspace{10mm}
\date{\today}

\begin{abstract}
Stenull and Lubensky [Phys.~Rev.~E
  {\bf 76}, 011706 (2007)] have argued that shear strain and tilt of the director
  relative to the layer normal are coupled in smectic elastomers and
  that the imposition of one necessarily leads to the development of
  the other. This means, in particular, that a Smectic-$A$ elastomer
  subjected to a simple shear will develop Smectic-$C$-like tilt of
  the director. Recently, Kramer and Finkelmann [arXiv:0708.2024, Phys.~Rev.~E
  {\bf 78}, 021704 (2008)]
  performed shear experiments on Smectic-$A$ elastomers using two different shear geometries.
  One of the experiments, which implements simple shear, produces clear evidence for the
  development of Smectic-$C$-like tilt.
  Here, we generalize a model for smectic elastomers introduced by
  Adams and Warner [Phys.~Rev.~E {\bf 71}, 021708 (2005)] and use it to study the magnitude of
  Sm$C$-like tilt under shear for the two geometries investigated by Kramer and Finkelmann.
  Using reasonable estimates of model parameters, we estimate the tilt angle for both geometries,
  and we compare our estimates to the experimental results.  The
  other shear geometry is problematic since it introduces additional
  in-plane compressions in a sheet-like sample, thus inducing
  instabilities that we discuss.
\end{abstract}


\maketitle

\section{Introduction}
\label{sec:introduction}

Smectic elastomers~\cite{WarnerTer2003} are rubbery materials with
the orientational properties of smectic liquid
crystals~\cite{deGennesProst93_Chandrasekhar92}. They possess a
plane-like, lamellar modulation of density in one direction. In the
smectic-$A$ (Sm$A$) phase, the Frank director $\nv$ describing
the average orientation of constituent mesogens is parallel to the
normal $\Nv$ of the smectic layers whereas in the smectic-$C$
(Sm$C$) phase, there is a non-zero tilt-angle $\Theta$ between
$\nv$ and $\Nv$.

Recently, there has been some controversy about whether shear strain
and tilt of the director relative to the layer normal are coupled in
smectic elastomers and whether the imposition of one necessarily
leads to the development of the other. Beautiful experiments by
Nishikawa and Finkelmann~\cite{nishikawa_finkelmann_99}, where a
Sm$A$ elastomer was subjected to extensional strain along the layer
normal, found a drastic decrease in Young's modulus at a threshold
strain of about $3\%$ accompanied by a rotation of $\Nv$ through an
angle $\varphi$ that sets in at the same threshold. Interpreting
their x-ray data, the authors concluded that there was no Sm$C$-like
order over the range of strains they probed, and that the reduction
in Young's modulus stems from a partial breakdown of smectic
layering. Recently, Adams and Warner (AW)~\cite{adams_warner_2005}
developed a model for Sm$A$ elastomers which assumes that $\nv$ and
$\Nv$ are rigidly locked such that $\Theta=0$. This model produces a
stress-strain curve and a curve for the rotation angle $\varphi$ of
$\Nv$ in full agreement with the experimental curves but without
needing to invoke a breakdown of smectic layering. Evidently,
because of the assumption $\Theta=0$, the AW model predicts no
Sm$C$-like order. More recently, Stenull and
Lubensky~\cite{StenLubSmA2007} argued that shear strain and
Sm$C$-like order are coupled and that the imposition of one
inevitably leads to the development of the other. They developed a
model based on Lagrangian elasticity that predicts, as does the AW
model, a stress-strain curve and a curve for $\varphi$ in full
agreement with the experiment. In contrast to the interpretation of
Nishikawa and Finkelmann of their data and to the central assumption
of the AW model, Ref.~\cite{StenLubSmA2007} found that the tilt
angle $\varphi_{\nv}$ of $\nv$ is not identical to that of the layer
normal, $\varphi$, implying that there is Sm$C$-like order with
non-zero $\Theta$ above the threshold strain. However, estimates of
$\varphi$ and $\varphi_{\nv}$ based on reasonable assumptions guided
by the available experimental data of the layer normal-director
coupling, turn out the same order of magnitude. The upshot is that
Ref.~\cite{StenLubSmA2007} predicts Sm$C$-like tilt above the
threshold strain but that the angle $\Theta$ is small. It is
entirely possible that $\Theta$ is smaller than the resolution of
the experiments by Nishikawa and Finkelmann. In this case, there is
no disagreement between the predictions of
Ref.~\cite{StenLubSmA2007} and the experimental data by Nishikawa
and Finkelmann.

The arguments of Ref.~\cite{StenLubSmA2007} imply in particular that
a Sm$A$ elastomer subjected to a shear in the plane containing $\Nv$
will develop Sm$C$-like tilt of the director. Very recently, Kramer
and Finkelmann (KF)~\cite{KramerFinkel2007,KramerFinkel2008}
performed corresponding shear experiments using two different shear
geometries. One geometry, which we refer to as tilt geometry,
imposes a shear that is accompanied
by an effective compression of the sample along the layer normal. In
this geometry, the elastomer ruptures at an imposed mechanical shear
angle $\phi$ of $13 \, \mbox{deg}$~\cite{KramerFinkel2008} or $14 \,
\mbox{deg}$~\cite{KramerFinkel2007}, and up to these values of
$\phi$ no Sm$C$-like tilt is detected within the accuracy of the
experiments of about $1 \, \mbox{deg}$, as was the case in the
earlier stretching experiments of Nishikawa and Finkelmann. The
other shear geometry, which we refer to as slider geometry, imposes
simple shear and can be used to probe values of $\phi$ exceeding $20
\, \mbox{deg}$. For this geometry, the KF x-ray data provides clear
evidence for the emergence of Sm$C$-like tilt.

In this communication, we generalize the model introduced by
AW~\cite{adams_warner_2005} and use it to study the magnitude of
Sm$C$-like tilt under shear for the two geometries investigated by KF.
Using reasonable estimates of model
parameters, we estimate the tilt angle $\Theta$, and we compare our estimates to the experimental results.

The outline of the remainder of this communication is as follows. In
Sec.~\ref{sec:extendingAW} we develop our model, which generalizes
the original AW model. In Sec.~\ref{sec:ExpGeomTiltAng} we describe
the two experimental setups that we consider. We discuss the
respective deformation tensors for these setups and calculate for
both setups the Sm$C$-tilt angle $\Theta$ as a function of the
imposed mechanical tilt angle $\phi$. We address why the
tilter apparatus is problematic for experiments where shear induces
 director tilt.  In Sec.~\ref{sec:discussion} we discuss our
findings and we make some concluding comments.  There are two
appendices. In App.~\ref{app:smallAngles}, we comment on analytical
calculations of $\Theta$ for small $\phi$. In App.~\ref{app:Euler},
we briefly discuss the Euler instability in the context of the
experiments by KF.

\section{Generalizing the Adams-Warner model}
\label{sec:extendingAW}
In this section we devise a theory for smectic elastomers based on
the {\em neo-classical} approach, developed originally by Warner and
Terentjev and coworkers~\cite{BlandonWar1994,WarnerTer2003} for
nematic elastomers, and the subsequent extension of the {\em
neo-classical} approach by AW to smectics.  The {\em neo-classical}
approach generalizes the classical theory of rubber
elasticity~\cite{treloar1975} to include the effects of
orientational anisotropy on the random walks of constituent polymer
links.  It treats large strains with the same ease as they are
treated in rubbers in the absence of orientational order, it
provides direct estimates of the magnitudes of elastic energies, it
is characterized by a small number of parameters, and it accounts
easily for incompressibility.

In the {\em neo-classical} approach, one formulates elastic energy
densities in terms of the Cauchy deformation tensor
$\tens{\Lambda}$, defined by $\Lambda_{ij} = \partial R_i/\partial
x_j$, where $\brm{R} (\brm{x})$ is the target space vector that
measures the position in the deformed medium of a mass point that
was at position $\brm{x}$ in the undeformed reference medium. In
this approach, a generic model elastic energy density for smectic
elastomers that allows for a relative tilt between the director and
the layer normal can be written in the form
\begin{align}
\label{eq:full_f}
f = f_{\text{trace}} + f_{\text{layer}} + f_{\text{tilt}} +
f_{\text{semi}} \, .
\end{align}
Here and in the following, incompressibility of the material is
assumed, i.e., the deformation tensor is subject to the constraint
$\det \tens{\Lambda} =1$. $f_{\text{trace}}$ is the usual trace
formula of the {\em neo-classical} model with $\mu$ the shear
modulus,
\begin{align}
f_{\text{trace}} = \textstyle{\frac{1}{2}} \mu \mbox{Tr} \left[
\tens{\Lambda} \, \tens{\ell}_0 \,  \tens{\Lambda}^T
\tens{\ell}^{-1} \right] \, ,
\end{align}
where $\tens{\ell}_0 = \tens{\delta} + (\ar-1) \nv_0 \nv_0$, with
$\nv_0$ describing the uniaxial direction before deformation, is the
so-called shape tensor describing the distribution of
conformations of polymeric chains before deformation and
$\tens{\ell}^{-1} = \tens{\delta} - (1-\ar^{-1}) \nv \nv$ is the
inverse shape tensor after deformation.  $\tens{\delta}$ denotes the
unit matrix, and $\ar$ denotes the anisotropy ratio of the uniaxial
Sm$A$ state. The contribution
\begin{align}
\label{layerCompressionEn}
f_{\text{layer}} = \frac{B}{2} \left[ \left( \frac{d}{d_0\, \cos
\Theta}\right)^2 -1 \right]^2
\end{align}
describes changes in the spacing of smectic layers with layer normal
\begin{align}
\Nv = \frac{\tens{\Lambda}^{-T}\Nv_0}{|\tens{\Lambda}^{-T}\Nv_0|}  \, ,
\end{align}
where $\Nv_0 = \nv_0$ is the layer normal before deformation and $B$
the layer compression modulus~\cite{footnoteCompressionTerm}. $d_0$ and $d$ are,
respectively, the layer spacing before and after deformation which are
related via
\begin{align}
\frac{d}{d_0} = \frac{1}{\big|\tens{\Lambda}^{-T}
\Nv_0 \big|} \, .
\end{align}
The tilt energy density,
\begin{align}
f_{\text{tilt}} =  \textstyle{\frac{1}{2}} \, \scp \sin^2 \Theta ,
\end{align}
incorporates into the model the preference for the director to be
parallel to the layer normal in the Sm$A$ phase. To study the Sm$C$
phase, we would have to include a term proportional to $\sin^4
\Theta$, which, however, is inconsequential for our current purposes.
Without the contribution $f_{\text{semi}}$, the model elastic energy
density~(\ref{eq:full_f}) is invariant with respect to simultaneous
rotations of the smectic layers, the nematic director, and
$\tens{\Lambda}$ in the target space.  To break this unphysical
invariance, we include the semisoft term \cite{WarnerTer2003}
\begin{align}
f_{\rm semi}&=  \textstyle{\frac{1}{2}} \mu \alpha \, {\rm
Tr}[(\tens{\delta} - \nv_0 \nv_0) \tens{\Lambda}^T \nv \nv
\tens{\Lambda}]   ,
\end{align}
where $\alpha$ is a dimensionless parameter.

The relation of the model presented here to the AW model is the following:
AW assume that the layer normal and the director are rigidly locked
such that the angle $\Theta$ is constrained to zero. Moreover, the
semi-soft term $f_{\text{semi}}$ is absent in the AW model.
Essentially, we retrieve the AW model from Eq.~(\ref{eq:full_f}) by
setting $\Theta = 0$ (or equivalently $\scp \rightarrow \infty$) and
$\alpha = 0$.

As mentioned above, one of the virtues of the {\em neo-classical}
approach is that it involves only a few parameters, and for most of
these there exist experimental estimates, which we will now review
briefly.  The shear modulus $\mu$ for rubbery materials is typically
of the order of $10^5 - 10^6 \, \mbox{Pa}$. A typical value for the
smectic layer compression modulus in smectic elastomers, as observed
in experiments with small strains along the layer normal, is
$B\sim10^7 \, \mbox{Pa}$, which is greater than the values in liquid
smectics. In previous experiments by the Freiburg group, the
anisotropy ratio was approximately
$\ar\approx1.1$~\cite{nishikawa_finkelmann_99}, and we adopt this
value for our arguments here. The value of $\scp$ can be estimated
from experiments by Brehmer, Zentel, Gieselmann, Germer and
Zungenmaier~\cite{brehmer&Co_2006} on smectic elastomers and by Archer
and Dierking~\cite{ArcherDie2005} on liquid smectics.  The former
experiment indicates that $\scp$ is of the order of $10^5 \,
\mbox{Pa}$ at room-temperature, and the latter experiment produces a
room-temperature value of the order of $10^6 \, \mbox{Pa}$. We are not
aware of any experimental data that allows us to estimate $\alpha$ for
smectic elastomers reliably. For nematic elastomers, it has been
estimated from the Fredericks effect~\cite{TerWarMeyYam1999} and from
the magnitude of the threshold to director rotation in response to
stretches applied perpendicular to the original
director~\cite{FinKunTerWar1997} that $\alpha \approx 0.06$ or $\alpha
\approx 0.1$, respectively. For our arguments here, we adopt the
latter value acknowledging that $\alpha$ may be considerably larger in
smectics than in nematics. As we will see in the following, our
findings do not depend sensitively on this assumption.  For the
deformations that we consider, $\alpha$ appears only in the
combination $\zeta = \scp + \alpha \mu$, that is, semi-softness simply
adds to the director-layer normal coupling.  Uncertainty in estimates
for $\zeta$ is expected stem mainly from the spread in estimates for
$\scp$. We account for this spread by discussing several values of
$\zeta$ or, more precisely, for several values of $\zeta/\mu$.

\section{Experimental shear geometries and tilt angles}
\label{sec:ExpGeomTiltAng}
\begin{figure}
\centerline{\includegraphics[width=6.0cm]{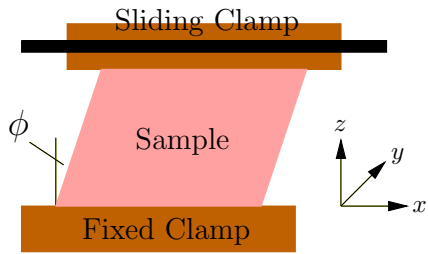}}
\caption{(Color online) Sketch of a slider apparatus where the upper
clamp slides on a horizontal bar such that the extension of the
sample in the $z$-direction remains fixed.}
\label{fig:sliderApparatus}
\end{figure}

\begin{figure}
\centerline{\includegraphics[width=5.5cm]{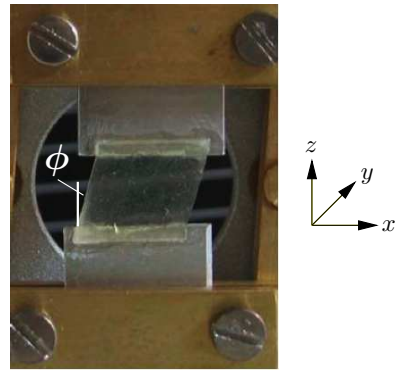}}
\caption{(Color online) One of the two tilt apparatuses used
  by KF. The photo has been taken
  from Ref.~\cite{KramerFinkel2007}.}
\label{fig:tiltApparatus}
\end{figure}

In this section we apply the model defined in
Sec.~\ref{sec:extendingAW} to study the behavior of Sm$A$ elastomers
in shear experiments. We consider two experimental setups. In the
first setup, which is perhaps the first that comes to mind from a
physicist's viewpoint, a simple shear is applied, i.e., a shear strain
in which the externally imposed displacements all lie in a single
direction.  Figure~\ref{fig:sliderApparatus} shows a sketch of such an
experiment, and we call the apparatus sketched in it the slider
apparatus.
The second setup, which is depicted in Fig.~\ref{fig:tiltApparatus},
is one in which opposing surfaces of the sample remain essentially
parallel but in which the height of the sample (its extension in the
$z$-direction) decreases upon shearing. Hence, the applied shear is,
strictly speaking, not just simple shear. In the
following we will refer to the this apparatus as a tilt apparatus.

\subsection{Slider apparatus}
\label{sec:sliderApparatus}
To make our arguments to follow as concrete and simple as possible,
we now choose a specific coordinate system. That is, we choose our
$z$ direction along the uniaxial direction $\nv_0$ of the unsheared
samples, $\nv_0 = (0, 0, 1)$, and we choose our $x$ direction as the
direction along which the sample is clamped, which is perpendicular
to $\nv_0$, cf.\ Fig.~\ref{fig:sliderApparatus}. With these
coordinates, the deformation tensor for both the slider apparatus
and the tilt apparatus is of the form
\begin{equation}
\label{formOfLambda}
\tens{\Lambda}= \left(
\begin{array}{ccc}
\Lambda_{xx} & 0 &  \Lambda_{xz} \\
0 & \Lambda_{yy} & 0 \\
0 & 0 & \Lambda_{zz}
\end{array}
\right) .
\end{equation}
As can be easily checked, for this type of deformation, the layer
normal lies along the
$z$-direction for the unsheared {\em and} for the sheared samples.
Thus, the tilt angle $\Theta$ between $\nv$ and $\Nv$ is identical
to the tilt angle between $\nv$ and the $z$-axis, and we can
parametrize the director as
\begin{align}
\nv = (\sin \Theta, 0, \cos \Theta) \, .
\end{align}
It is worth noting that a deformation tensor of the form shown in
Eq.~(\ref{formOfLambda}) leads to a particularly simple expression for
the semi-soft contribution to $f$, $f_{\rm semi} = (1/2) \alpha \mu
\Lambda_{xx}^2\sin^2 \Theta$, which combines with the tilt energy
density to $f_{\text{tilt}} + f_{\rm semi} = (1/2) \zeta (
  \Lambda_{xx}) \sin^2 \Theta$. Thus, as indicated above, our model
depends for the experimental geometries under consideration on
$\alpha$ and $\scp$ through a single effective parameter, viz.\
$\zeta( \Lambda_{xx})= \scp + \Lambda_{xx}^2\alpha \mu$.

From the photos of the experimental samples provided in
Ref.~\cite{KramerFinkel2007}, see Figs.~\ref{fig:sliderApparatus}
and \ref{fig:tiltApparatus}, it appears as if $\Lambda_{xx}$ does
not deviate significantly from 1. In the following, we set
$\Lambda_{xx} = 1$ for simplicity~\cite{footnoteLambdaXX}. In this
event $\zeta ( \Lambda_{xx})= \zeta \equiv \scp + \alpha \mu$. In
the slider apparatus, the extent of the sample in the $z$-direction
is fixed, and hence $\Lambda_{zz} = 1$. The incompressibility
constraint $\det \tens{\Lambda} =1$ thus mandates that $\Lambda_{yy}
= 1$. The remaining nonzero component of the deformation tensor is
entirely determined by the externally imposed shear, $\Lambda_{xz} =
\tan \phi$, where $\phi$ is the mechanical tilt angle of the sample,
see Fig.~\ref{fig:sliderApparatus}. The only remaining degree of
freedom in the problem is, therefore, the angle $\Theta$.

\begin{figure}
\centerline{\includegraphics[width=7.4cm]{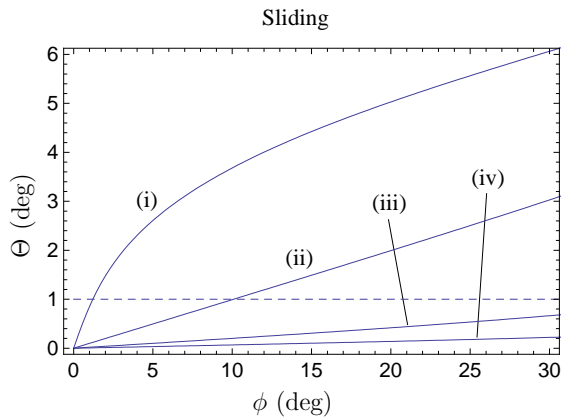}}
\caption{The tilt angle $\Theta$ between the layer normal and the
  director as a function of the mechanical tilt angle $\phi$ in the
  slider apparatus for (i) $\zeta/\mu = 0.1$, (ii) $\zeta/\mu = 1$,
  (iii) $\zeta/\mu = 5$, and (iv) $\zeta/\mu = 20$. The dashed line
  corresponds to the resolution for $\Theta$ in the experiments by
  KF.}
 \label{fig:ThetaSliderApparatus}
\end{figure}

To calculate $\Theta$ as a function of $\phi$, we insert the just
discussed deformation tensor into $f$ and then minimize $f$ over
$\Theta$ holding $\phi$ fixed. For $\phi$ small, this can be done
analytically by expanding $f$ to harmonic order in $\Theta$ and by
then solving the resulting linear equation of state for equilibrium
value of $\Theta$. This type of analysis is presented in the appendix.
To provide reliable predictions for larger shears, one has to refrain
from expanding in $\Theta$ and use numerical methods instead. To this
end, we minimize $f$ numerically assuming, based on what we discussed
at the end of Sec.~\ref{sec:extendingAW}, that $B/\mu = 10$ and
$\zeta/\mu \in \{0.1, 1, 5 ,20 \}$. For this minimization, we use
Mathematica's FindMinimum routine.
Figure~\ref{fig:ThetaSliderApparatus} shows the resulting curves for
$\Theta$ as a function of $\phi$. The dashed horizontal line in
Fig.~\ref{fig:ThetaSliderApparatus} is a guide to the eye; it
corresponds to the angle resolution in the KF experiments which was
about $1$ degree~\cite{kramerPrivate}.

For the slider geometry, the KF data produces clear evidence for the development of Sm$C$-like tilt under simple shear, and our theoretical estimates agree well with the experimental data.
However, given that thus far only two data points are available for $\phi > 0$, it cannot be judged reliably whether our theoretical curve could be fitted to an experimental curve with more data points. It is encouraging, though, that our curve for $\zeta/\mu = 0.1$ agrees with the available experimental points within their errors.

\subsection{Tilt apparatus}

Now we turn to the tilt geometry. The essential difference between
the slider and the tilt geometry is that in the former the height of
the sample $L_z (\phi)$ remains constant, $L_z (\phi) = L_z$ ($d_0$
times the number of smectic layers) whereas in an ideal
tilter, $L_z (\phi)$ is not constant but rather decreases as $\phi$
increases, $L_z (\phi) = L_z \cos \phi$. This difference has far
reaching consequences. In the slider, the sample-height
remains larger than the modified natural
height of the sample created by the director tilt, $L_z \cos
\Theta$. In other words, the sample is under effective tension. In
the tilter, on the other hand, the sample-height $L_z \cos \phi$ can
be smaller than $L_z \cos \Theta$, i.e., the sample can be under
effective compression. For a sample under effective $zz$-compression, one has to worry, experimentally and theoretically, about all sorts of complications. Most notable are
perhaps buckling and wrinkling.

The theory of buckling of elastic sheets is well established~\cite{Landau-elas}, and we now briefly comment on buckling in the context of the tilt geometry. As mentioned above, a  sample clamped into the tilt apparatus can develop a buckling instability, if the height of the sample imposed by tilt, $L_z \cos
\phi$, is smaller than the natural height of the sample created by the director tilt, $L_z \cos \Theta$. As we will see below, our results for $\Theta$ as a function of $\phi$ imply that $\cos \phi < \cos \Theta$, i.e., buckling is possible. Alternatively, this can be seen by calculating the engineering stress
$\sigma^{\text{eng}}_{zz} = \partial f/\partial \Lambda_{zz}$, which can be done using the numerical approach outline above. $\sigma^{\text{eng}}_{zz}$ is positive for $\phi>0$ in the slider
apparatus, whereas it turns out being negative for $\phi>0$ in the tilt apparatus. Thus, the sample is effectively under
tension in the slider apparatus whereas it is effectively under
compression in the tilt apparatus, opening the possibility for
buckling in the $y$-direction in the latter. The angle $\phi^c$ at
which buckling sets in is expected to be comparable to that for the
well known Euler Strut instability~\cite{Landau-elas},
\begin{align}
\label{EulerAngle}
\phi^c \approx \phi^{\text{Euler}} = \arccos \left[ 1 -  \left( \frac{2 \, \pi\, L_y}{3\, L_z} \right)^2 \right],
\end{align}
where $L_y$ is the thickness of the sample in the $y$-direction, and
where clamped (rather than hinged) boundary conditions are assumed. A
brief derivation of Eq.~(\ref{EulerAngle}) is given in
App.~\ref{app:Euler}. In the experiments of KF, $L_z = 5.0 \,
\mbox{mm}$ and $L_y = 0.45 \, \mbox{mm}$~\cite{kramerPrivate}, which
leads to $\phi^{\text{Euler}} \approx 15 \, \mbox{deg}$.
The experimental samples buckle immediately before they rupture~\cite{kramerPrivate} at $\phi = 13 \, \mbox{deg}$ or $\phi = 14 \, \mbox{deg}$, which is very close to our estimate for $\phi^c$. The observation that buckling occurs immediately before rupturing might suggest that the former actually triggers the latter.

A detailed analysis of sample-wrinkling in the tilt geometry is
beyond the scope of the present paper. However, a few comments about
wrinkling are in order.  The wavelength of wrinkling is expected to
be much shorter than that of buckling. Thus, wrinkling can be harder
to detect by visual inspection of an experimental sample than
buckling, and there is a risk that it remains unnoticed. The main
problem is, however, that wrinkling can act as to effectively reduce
the mechanical shear in the sample. When designing a tilt
experiment, one thus has be very careful to avoid wrinkling. If not,
the effects of wrinkling can bias the data, and there is the risk to
underestimate the Sm$C$-tilt significantly.

\begin{figure}
\centerline{\includegraphics[width=7.4cm]{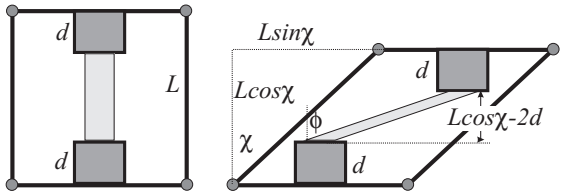}}
\caption{Schematic of shear and compression arising in a non-ideal
tilter.}
 \label{fig:tiltSchematic}
\end{figure}

Our calculation of the the Sm$C$-tilt will be for an ideal tilter.
Figure~\ref{fig:tiltSchematic} shows the tilter of
Fig.~\ref{fig:tiltApparatus} schematically in a non-tilted and a
tilted configuration to make clear that shear and compression are
complex in the non-ideal case because the frame-axles allowing angle
change are off-set from the corners of the sample.
Using elementary trigonometry, one can deduce for the shears and the angle $\phi$
\begin{eqnarray}
\lambda_{xz} &=& \frac{L}{L-2d} \sin
\chi\, , \;\;\;\;\; \lambda_{zz} = \frac{L\cos \chi -2 d}{L-2d}\, ,
\label{eq:non-ideal-shears}\\
\tan\phi &=& \frac{\tan \chi}{1 - (2d/L) \sec \chi} \, ,
\label{eq:non-ideal-angle}
\end{eqnarray}
In particular, the relation between the shear angle $\phi$ and the
deformation components $\lambda_{zz}$ and $\lambda_{xz}$ is not
simple.

\begin{figure}
\centerline{\includegraphics[width=7.5cm]{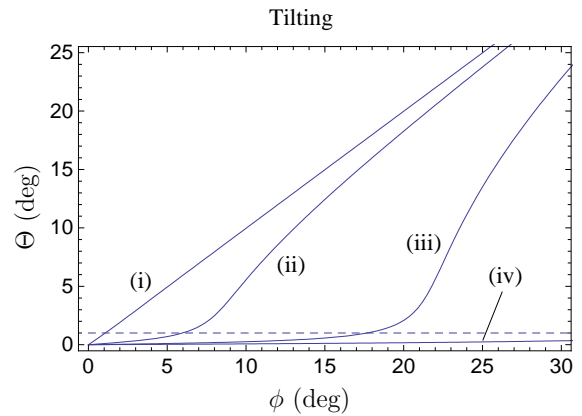}}
\caption{The tilt angle $\Theta$ between the layer normal and the
  director as a function of the mechanical tilt angle $\phi$ in the
  tilt apparatus for (i) $\zeta/\mu = 0.1$, (ii) $\zeta/\mu = 1$,
  (iii) $\zeta/\mu = 5$, and (iv) $\zeta/\mu = 20$. The dashed line
  corresponds to the resolution for $\Theta$ in the experiments by
  KF.}
\label{fig:ThetaTiltApparatus}
\end{figure}

Returning to an ideal tilter, $d = 0$, $\chi \equiv \phi$, we
calculate the Sm$C$-tilt as a function of the applied mechanical
shear, suppressing buckling and wrinkling. Then, the deformation
tensor is of the same form~(\ref{formOfLambda}) as for the slider
apparatus, but with the essential difference that here $\Lambda_{xz}
= \sin \phi$ and $\Lambda_{zz} = \cos \phi$. As we did for the
slider apparatus, we assume $\Lambda_{xx} = 1$. The
incompressibility constraint then implies that $\Lambda_{yy} =
1/\cos \phi$, leaving the angle $\Theta$ as the only degree of
freedom. We calculate $\Theta$ as a function of $\phi$ in exactly
the same way as above, i.e., we minimize $f$ numerically assuming
the values of the model parameters discussed in
Sec.~\ref{sec:extendingAW}. The resulting curves are shown in
Fig.~\ref{fig:ThetaTiltApparatus}.
The dip in Fig.~\ref{fig:ThetaTiltApparatus} occurs because of a
competition between $\zeta$ which would prefer $\Theta = 0$ and $B$
which would prefer $\Theta = \phi$. If $\zeta>B$ then it is possible
to get curves where $\Theta$ stays close to $0$. The dashed
horizontal line in Fig.~\ref{fig:ThetaTiltApparatus} indicates the
angle resolution of about $1$ degree of the KF
experiments~\cite{kramerPrivate}.

As mentioned in Sec.~\ref{sec:sliderApparatus}, the data points for
the slider are compatible with $\zeta/\mu \approx 0.1$. The
elastomers used by KF in the slider and the tilter are identical, and, therefore, curve (i)
of Fig.~\ref{fig:ThetaTiltApparatus} should describe the Sm$C$-tilt
if the tilter used by KF were ideal or nearly so. Note from
Fig.~\ref{fig:ThetaTiltApparatus}, however, that this implies that
the Sm$C$-tilt for an ideal tilter at $\phi =  13$ or $14  \,
\mbox{deg}$ should be significantly larger than the experimental
resolution, which is incompatible with the findings of
Refs.~\cite{KramerFinkel2007,KramerFinkel2008}.

Since any effective compression of the sample along the layer
normal, as in a tilter, promotes rather than hampers director
rotation, we obtain a Sm$C$-tilt at a given $\phi$ in a tilter if it
occurs at the same angle in the slider. It is likely that the
failure of KF to observe director rotation in their tilter is due to
mechanical instability  that  effectively reduces the mechanical
shear of the sample and thus leads to a systematic suppression of
the Sm$C$-tilt in the tilt geometry.

\section{Discussion and concluding remarks}
\label{sec:discussion}
In summary, we have developed a {\em neo-classical} model for
smectic elastomers, and we used this model to study Sm$A$ elastomers
under shear in the plane containing the director and the layer
normal. In particular, we investigated the tilt angle $\Theta$
between the layer normal and the director for two different
experimental setups as a function of the mechanical tilt angle
$\phi$ measuring the imposed shear.

Our model builds upon the {\em neo-classical} model for smectic
elastomers by AW. In their original work, AW chose not to consider
the possibility of Sm$C$ order in their theory: they forced the
layer normal and the nematic director to be parallel. In our present
theory, the fundamental tenet is different in that the nematic
director $\nv$ and the deformation tensor $\tens{\Lambda}$  are
independent quantities that must be allowed to seek their
equilibrium in the presence of imposed strains or stresses.
The layer normal $\Nv$ is determined entirely by
$\tens{\Lambda}$. Thus, $\nv$ and $\Nv$ are independent variables
that rotate to minimize the free energy - they are not locked
together.

In the present paper, we have for simplicity focused on idealized
monodomain samples. Boundary conditions in real experiments prevent
this simple scenario and produce a microstructure structure of the
type that AW discuss in Ref.~\cite{adams_warner_2005} based on their
original model that prohibits Sm$C$ ordering. Our current theory,
which admits Sm$C$ ordering, predicts the same type of microstructure;
stretching will produce a polydomain layer structure just as the AW
theory does. In particular, the existence of Sm$C$ ordering does not
contradict the appearance of optical cloudiness. To avoid undue
repetition, we refrain discussing this here in more detail and refer
the reader to Ref.~\cite{adams_warner_2005}.

The main finding of our present work is that the tilt angle $\Theta$
between $\nv$ and $\Nv$ is non-zero if a shear in the plane
containing $\Nv$ is imposed. This angle depends on material
parameters, the experimental setup, and the magnitude of the imposed
shear.  Figures~\ref{fig:ThetaSliderApparatus} and
\ref{fig:ThetaTiltApparatus} depict our results for the slider
apparatus and the tilt apparatus used by KF.  As expected, in both
setups the angle $\Theta$ decreases as the value of the parameter
$\zeta/\mu$ increases, at fixed $\phi$.  This is because the
director is becoming more strongly anchored to the layer normal. If
one keeps $\zeta/\mu$ fixed and varies $B/\mu$ instead, then for
$B/\mu\to \infty$ $\Theta$ approaches $\phi$ because in this limit
$f_{\text{layer}}$ locks $\Theta$ to $\Theta = \phi$. As $B/\mu$ is
decreased then $\Theta$ decreases at fixed $\phi$, because the
$f_\mathrm{tilt}$ term starts to compete with the $f_\mathrm{layer}$
term. In the opposite limit of very small $B/\mu$, $\Theta$ is
locked to $\Theta = 0$ by virtue of the tilt and semi-soft
contributions to $f$. We refrain to show the corresponding curves,
which are similar to those depicted in
Figs.~\ref{fig:ThetaSliderApparatus} and
\ref{fig:ThetaTiltApparatus}, to save space.  The main difference
between our results for the two setups is that for $B>\zeta$ and
large $\phi$, say $40$ degrees or so, $\Theta$ is roughly one order
of magnitude smaller than $\phi$ in the slider apparatus, whereas it
is of the order of $\phi$ in the tilt apparatus.

In closing, we would like to stress once more that the experiment
using the slider geometry produces clear evidence of the development
of Sm$C$-like tilt under shear as predicted originally in
Ref.~\cite{StenLubSmA2007}. Our theory developed here allowed us to
understand in some detail how the magnitude of this tilt depends on
the shear geometry and on model/material parameters, such as
$\zeta$, which can be reasonably estimated from nematic analogues.
Given these estimates, the resulting estimates for the tilt angle
vary over a range from being small enough to be essentially
unobservable with the X-ray equipment used by the Freiburg group to
exceeding the experimental resolution. While the latter is
consistent with the shear experiments in the slider geometry, the
former is consistent with the stretching experiments by Nishikawa
and Finkelmann in which no Sm$C$-like tilt was detected. We believe
that the reason for the discrepancy between the rotations in the
slider and tilter apparatus lies in a possible wrinkling of the
samples in the tilt apparatus used by KF. Mechanical instability
leads to an effective reduction of the mechanical shear of the
sample causing a systematic underestimation of the Sm$C$-like tilt
in the tilt geometry.

Another experimental approach that could be used, at least in principle, to
investigate $\Theta$ is to measure changes in the layer spacing, as
was done by Nishikawa and Finkelmann~\cite{nishikawa_finkelmann_99}.
It should be noted, however, that in this approach one measures $\cos
\Theta$ rather than $\Theta$ directly. For small values of Sm$C$
tilt, say $\Theta = 1$ degree, $\cos \Theta \approx
0.9998$ which is practically indistinguishable from the $\cos \Theta
=1$ pertaining to Sm$A$ order.

For a refined comparison between experiment and theory it would be
useful to critically evaluate and potentially improve the design of
the tilt apparatuses regarding sample-wrinkling. Also, it would be
interesting to perform shear and stretching experiments with an
angle resolution better than $1\, \mbox{deg}$. We hope that our work
stimulates the interest for such experiments.

\begin{acknowledgments}
  This work was supported in part by the National Science Foundation under
  grants No.\ DMR 0404670 and No.\ DMR 0520020 (NSF MRSEC) (T.C.L.). We are grateful to D.~Kramer and H.~Finkelmann for discussing with us some of the details of their
  experiments.
\end{acknowledgments}

\appendix
\section{Analytical consideration for small angles}
\label{app:smallAngles}
For arbitrary Sm$C$-tilt angle $\Theta$, the total elastic energy
density $f$ is, for the deformations discussed in
Sec.~\ref{sec:ExpGeomTiltAng}, a fairly complicated conglomerate of
trigonometric functions (and powers thereof) of $\Theta$ and $\phi$.
Thus, we resorted in Sec.~\ref{sec:ExpGeomTiltAng} to a numerical
approach to determine the equilibrium value of $\Theta$ over a wide
range of the imposed mechanical tilt angle $\phi$. Here, we focus on
the regime of $\phi$ where $\Theta$ is small, such that it is
justified to expand $f$ to harmonic order in $\Theta$. In this case,
the equations of state are linear in $\Theta$, of course, and are
therefore readily solved. For the slider apparatus we obtain
\begin{align}
\label{ThetaAnaSlider}
\Theta = \frac{(\ar-1)\ar \mu \tan \phi}{\ar \zeta+ \mu (1-\ar) [1-\ar + \ar \tan^2 \phi]} \, ,
\end{align}
and for the tilt apparatus we get
\begin{align}
\label{ThetaAnaTilt}
\Theta = \frac{2^{-1} (\ar-1)\ar \mu \sin 2 \phi}{\ar \zeta - \ar B \sin^2 2\phi + \mu (1-\ar) [1-\ar \cos 2 \phi]} \, .
\end{align}
Equations~(\ref{ThetaAnaSlider}) and (\ref{ThetaAnaTilt}) imply, that
the Sm$C$-tilt in both setups is identical for small $\phi$,
\begin{align}
\label{ThetaAnaExpanded}
\Theta = \frac{(\ar-1)\ar \mu \,  \phi}{\ar \zeta + \mu (1-\ar)^2} + O(\phi^3)\, .
\end{align}
This is consistent with Figs.~\ref{fig:ThetaSliderApparatus} and
\ref{fig:ThetaTiltApparatus}, where the initial slopes are identical
for identical values of $\zeta$ (though this is perhaps somewhat hard
to see because the ordinate-scales are different in the two figures).

An interesting related question, which we have not addressed in the
main text because there is currently no experimental data available,
is that of the stress that is caused by the imposed shear strains.
From the above, it is straightforward to calculate this stress for
small $\phi$. Inserting Eq.~(\ref{ThetaAnaExpanded}) into the
aforementioned harmonic (in $\Theta$) total elastic energy density
provides us with an effective $f$ in terms of $\phi$. From this
effective $f$, we readily extract the engineering or first
Piola-Kirchhoff stress as
\begin{align}
\label{engineeringStress}
\sigma^{\text{eng}}_{xz} = \frac{\partial f}{\partial \Lambda_{xz}} = \frac{\ar^2 \mu  \zeta \, \phi}{\ar \zeta + \mu (1-\ar)^2} + O(\phi^3)\, .
\end{align}
Equation~(\ref{engineeringStress}) highlights a problem that arises
when the tilt and the semi-soft contributions $f_{\text{tilt}}$ and
$f_{\text{semi}}$ to the total elastic energy density are missing, and
one allows $\brm{n}$ and $\tens{\Lambda}$ to be independent
quantities. Setting $\zeta =0$ in Eq.~(\ref{engineeringStress}) leads
to zero stress for non-zero $\phi$, i.e., this truncated model
predicts soft elasticity. This soft elasticity, however, is not
compatible with Sm$A$ elastomers crosslinked in the Sm$A$ phase, such
as the experimental samples of
Refs.~\cite{nishikawa_finkelmann_99,KramerFinkel2007}, where the
anisotropy direction is permanently frozen into the system.

\section{Euler instability in the tilt apparatus}
\label{app:Euler}
In this appendix, we give a brief derivation of
Eq.~(\ref{EulerAngle}). We are interested primarily in a rough
estimate for $\phi^c$, and therefore, for simplicity, we assume that
we can ignore the effects of smectic layering. In the following, we
employ, for convenience, the Lagrangian formulation of elasticity
theory.

In the Lagrangian formulation, the elastic energy density of a thin
elastomeric film with thickness $L_y$ and height $L_z$ that is
compressed along the $z$-direction can be written as
\begin{align}
\label{fLagrangian}
f_{\text{film}} = \textstyle{\frac{1}{2}} \, \kappa \left(  \partial^2_z u_y\right)^2 +  \textstyle{\frac{1}{2}} \, Y_{2d} u_{zz}^2\, ,
\end{align}
where the choice of coordinates is the same as depicted in
Fig.~\ref{fig:tiltApparatus}. $u_y$ is the $y$-component of the
elastic displacement $\brm{u} = \brm{R} - \brm{x}$, and $u_{zz}^2$ is
the $zz$-component of the Cauchy-Saint-Venant strain tensor, $u_{ij} =
\frac{1}{2} (\partial_i u_j + \partial_j u_i + \partial_i
u_k\partial_j u_k)$. $\kappa$ and $Y_{2d}$ are, respectively, the
bending modulus and Young's modulus of the film, which are given in
the incompressible limit by $\kappa = (\mu/3) L_y^3$ and $Y_{2d} = 3
\mu L_y$~\cite{Landau-elas}. At leading order, $\partial_z u_z = -
|\delta L_z /L_z|$, where we have used that the height change $\delta
L_z$ is negative when the sample is effectively compressed as in the
tilt apparatus. This leads to
\begin{align}
\label{uzz}
u_{zz} = - |\delta L_z /L_z| + \textstyle{\frac{1}{2}} \, (\partial_z u_y )^2  ,
\end{align}
when we concentrate on the parts of $u_{zz}$ that are most important
with respect to buckling in the $y$-direction. Next, we substitute the
strain~(\ref{uzz}) into Eq.~(\ref{fLagrangian}) and switch to Fourier
space. To leading order in the elastic displacement, this produces
\begin{align}
\label{fFourier}
\tilde{f}_{\text{film}} =  \frac{1}{2}\left\{ \kappa \, q_z^2  - Y_{2d} \, \left|  \frac{\delta L_z}{L_z} \right| \right\} q_z^2 \, \tilde{u}_y (\brm{q}) \tilde{u}_y ( - \brm{q}) \, ,
\end{align}
where $\brm{q}$ is the wavevector conjugate to $\brm{x}$,
$\tilde{u}_y$ is the Fourier transform of $u_y$, and so on.
Equation~(\ref{fFourier}) makes it transparent that buckling occurs
for
\begin{align}
\label{deltaLoverL}
\left|  \frac{\delta L_z}{L_z} \right| = \frac{\kappa}{Y_{2d}} \, q_z^2= \frac{L_y^2}{9} \, q_z^2 \, .
\end{align}
The smallest value of $q_z$ is determined by the specifics of the
boundary conditions. From Fig.~\ref{fig:tiltApparatus} it appears as
if the sample of KF is clamped such that it prefers to stay parallel
to the clamps in their immediate vicinity. In this case, the smallest
value of $q_z$ is $q_z = 2\pi/L_z$. For hinged boundary conditions, in
comparison, its smallest value would be $q_z = \pi/L_z$.  Using the
former value and exploiting that, approximately, $|\delta L_z /L_z| =
1 - \cos \phi$, we obtain the estimate~(\ref{EulerAngle}) for the
onset of the Euler instability in the tilt apparatus.


\begin{thebibliography}{}
\bibitem{footnoteCompressionTerm} Note that the layer compression term
  formulated by AW is of the form $[d/d_0 -1]^2$ whereas setting
  $\Theta =0$ in Eq.~(\ref{layerCompressionEn}) produces $[(d/d_0)^2
  -1]^2$. This form is chosen for consistency with the
    Chen-Lubensky model of smectics. This term can be factorized into
    $[(d/d_0) -1]^2[(d/d_0) +1]^2\approx 4 [d/d_0 -1]^2$ for $d \sim
    d_0$. Thus the difference is merely factor of $4$ in the
    definition of the smectic modulus $B$ in the two representations.
\bibitem{footnoteLambdaXX}
We checked the validity of this assumption by allowing
$\Lambda_{xx}$ to relax to its equilibrium value, i.e., we repeated
our analysis using the $\Lambda_{xx}$ that minimizes $f$ rather than
$\Lambda_{xx} =1$. The results obtained by using this more involved
approach do not differ significantly from those obtained  our
simplified approach with $\Lambda_{xx} =1$.
\end{thebibliography}
\end{document}